\documentclass[prb,twocolumn]{revtex4-1}

\usepackage{amsmath}  
\usepackage{amsfonts} 
\usepackage{graphicx} 
\pdfoutput=1
\begin{document}

\title{Physics in the information age: qualitative methods (with examples from quantum mechanics).}

\author{V. G. Karpov}
\email{victor.karpov@utoledo.edu}
\affiliation{Department of Physics and Astronomy, University of Toledo, Toledo, OH 43606, USA}
\author{Maria Patmiou}
\email{maria.patmiou@rockets.utoledo.edu} 
\affiliation{Department of Physics and Astronomy, University of Toledo, Toledo, OH 43606, USA}

\date{\today}

\begin{abstract}
The traditional pedagogical paradigm in physics is based on a deductive approach. However, with the recent advances in information technology, we are facing a dramatic increase in the amount of readily available information; hence, the ability to memorize the material and provide rigorous derivations lacks significance. Our success in navigating the current ``sea" of information depends increasingly on our skills in  pattern recognition and prompt qualitative analysis. Inductive learning (using examples and intuition-based) is most suitable for the development of such skills. This needed change in our pedagogical paradigm remains yet to be addressed in physics curricula. We propose that incorporating inductive elements in teaching -by infusing qualitative methods - will better prepare us to deal with the new information landscape. These methods bring the learning experience closer to the realities of active research. As an example, we are presenting a compendium for teaching  qualitative methods in quantum mechanics, a traditionally non-intuitive subject.

\end{abstract}

\maketitle

\section{Introduction}
The traditional teaching approach in physics is deductive. After ascertaining that the necessary mathematical background has been acquired, the students are given a set of observations and principles to be used as a template in the future scientific adventures. The deductive approach is typical of the major physics textbooks and syllabi, and many instructors would agree that it is most economical and convincing in presenting the content of physical sciences. Its underlaying principles constitute a necessary part of a physicist's toolkit.

Inductive methods that would include heuristic arguments, guesswork and approximations are not taught as part of a regular physics curriculum. At the same time, it has long been known \cite{howpeoplelearn,howstudentslearn} and described in various forms (e.g.``The mind is not a vessel to be filled, but a fire to be kindled." Plutarch), that the dominance of inductive learning in scientific explorations matches our brain functionality. Indeed, as more and more evidenced from the modern understanding of the brain,  that functionality is based on the processes of pattern recognition and neural network training \cite{haykin,anderson}.

In reality, students usually become aware of inductive reasoning after they graduate, ‘from the horse’s mouth’ through their research, guided by experienced colleagues who -in turn- have been taught by their advisers. That way of learning underlies the existence of scientific schools of thought, per se, where junior participants can see “how it is done” in real-life physics, practice similar approaches, and develop a unique intuition based on properly built neural networks. (As one of many  examples, N. Bohr created a school of thought with the concerted effort of many who joined his seminar and adopted his thinking in the Institute of Theoretical Physics in Copenhagen; famous scientific schools include those by L. Landau, L. Euler, and many others).

The ongoing IT revolution has  accentuated the lack of scientific ``intuition" by making a tremendous amount of information available to us at an instant (through various networks and databases) while challenging the mind to recognize its rendered patterns. In particular, recent AI developments address the problem of pattern recognition in massive databases. And yet that necessity is insufficiently addressed in the typical physics curricula. The current teaching methods that continue the tradition of deductive learning need to be  augmented so as to better prepare students to deal with such challenges \cite{tofig}.

\section{Qualitative methods}\label{intro}

Along the lines of inductive learning in general and in physics in particular, the practicalities of inductive learning are significantly based on multiple examples (`patterns') analyzed by various qualitative methods. While approximate, these methods remain rigorous (we need not to confuse rigor with exactness). They are fast and intuitive enough to allow efficient pattern recognition, thus leading to the proper development of individual neural networks. That in turn, greatly facilitates the processes of recognition, classification and prediction.

It is not easy to answer the question of why qualitative methods are not represented in regular physics curricula \cite{chi,leonard,romer,haber,smith}. A list of possible factors includes:  i)historical details of the physics curricula development; ii)lack of relevant textbooks and expertise (very few textbooks on qualitative methods have been published as of now \cite{martynyuk,villaggio,krainov,caconi,migdal}); iii)underestimation of the importance of qualitative methods; iv)lack of interest in writing the corresponding textbooks (for example, the outstanding textbook by A. B. Migdal \cite{migdal} targets mostly professionals capable of understanding high enough level of presentation).

The qualitative or semi-quantitative “back of envelope” calculations are quite impressive for laymen due to their deceptive simplicity; it is less known that mastering such techniques requires special training and professional guidance. Furthermore, it requires a rather deep understanding of the subject matter, that allows one to neglect multiple features as insignificant and thus reduce the problem to its most simple representation.

Moving towards the substantive part of this paper, we note that multiple examples of qualitative methods are well known in various subject areas, such as the physics of fluids, statistical physics, the theory of disordered systems, etc. where exact solutions are impossible or useless. An important lesson, learned  from that multitude of applications, is that the order-of-magnitude approach provides results that are generally good to the accuracy of insignificant numerical factors, such as, say, 0.7, or 2.1, etc., as compared to the exact results (when available)\cite{karpov2015}. In the meantime, it is more intuitive, fast and economical, and avoids unnecessary mathematical complications.

In qualitative method examples, the equality sign $=$ can be replaced with the `equal in-the-order-of magnitude' sign ``$\sim$", meaning `to the accuracy of a numerical factor'. For example, the area of a circle of diameter $d$ is estimated as $A\sim d^2$, instead of the exact $A=\pi d^2/4$, different by a numerical factor of $\pi /4\approx 0.75$. (Of course, $A\sim d^2$ is numerically a better approximation than $A\sim r^2$ -where $r$ is the radius of a circle- in spite of them having the same dimensionality.) Below, we will compare, at some points, the `order-of-magnitude' and exact results.

In addition, the `order-of-magnitude' analyses often use intuitive estimates for derivatives and integrals, such as $\partial y/\partial x\sim \Delta y/\Delta x$ and $\int _a^{a+\Delta X}y(x)dx\sim y(a)\Delta x$ where $\Delta y$ and $\Delta x$ are the intervals for variables $x$ and $y$. (These approximations apply to smooth functions as opposed to i.e. strongly oscillating ones.) The approximations also generate additional numerical uncertainties. For example, assuming $y=x^2$, the latter estimates yield, $\partial y/\partial x\sim x^2/x=x$ and $\int _0^xydx\sim yx=x^3$, missing the numerical factors of 2 and 1/3 respectively, which are insignificant in the sense that they are not affecting the orders of magnitude.

The observed insignificance of numerical factors has found a commonly shared conceptual explanation (folklore attributed to Einstein): since numerical uncertainties at different steps in complex processes are mutually independent, they tend to cancel each other similarly to random displacements in a diffusion process (i. e. `errors propagate by diffusion'). For example, the coefficients of 2 and 1/3 in the above cases, can combined as 2/3, which is even closer to unity. There are very few known exceptions where such cancelation does not take place. This amazing observation is fully consistent with the fact that the laws of physics typically do not include any significant numerical multipliers when written in a `natural', way, e.g. in Gaussian units (as opposed to SI where such multipliers are enforced).

Quantum mechanics (QM) is a subject particularly suited for the application of qualitative methods because of the lack of corresponding ``everyday experiences" (we cannot visualize wavefunctions or operators). What follows is a number of examples of how qualitative results can be derived without copious mathematical work, but nevertheless reflect the basic principles taught in both college and graduate level quantum mechanics courses. These examples represent a compendium of ``tricks of the trade", methods used by many practicing physicists. Emphasis is now given on the development of -among others- qualitative understanding, simplified models, and order-of-magnitude estimates. The material below is aimed at a level lower than that of the famous Migdal's textbook \cite{migdal} and is accessible to a variety of college and grad students in science and engineering, as guided by qualified instructors.


\section{Dimensional Estimates in QM}\label{sec:DE}
\subsection{Dimensions of physical quantities}\label{sec:dpq}

Dimensional estimates are based on combining essential physical quantities to form expressions having the right dimensions \cite{bridgman,gibbings,gommes,robinett}. This is more systematically achieved when the dimensions of both the
underlying and estimated quantities are expressed in terms of {\it elemental} dimensions first introduced by the Gaussian system of units; these are: length
$[L]$, mass $[M]$, and time $[T]$.

The reasoning behind dimensional analysis is that -while not very often emphasized in college physics- the simple truth is that however complex a phenomenon that we study may be, {\it all} we can measure, is one of the above quantities (by reading scales, or measuring voltmeter needle rotations, or time lapsed). A related consideration is that our human design is such that we perceive and act through the eyes of classical mechanics, with our senses tuned to velocities, weights, forces, times, and their various combinations (it was natural then that classical mechanics turned
out to be historically the first part of physics developed).

Many quantum problems involve electric interactions. Therefore, we  will start our dimensional analysis with a law that is the cornerstone of electricity, Coulomb's law \cite{martinez,williams}, which establishes the {\it force} (mechanical quantity) between two charges at certain distance $r$. [We will write this law -as well as the subsequent material in this article- using the Gaussian (or cgs, with c=centimeter, g= gram, and s = second) system of units. While the SI system is more widespread, the cgs remains more suitable for making transparent the essence of the relationships between physical quantities \cite{sivukhin,littlejohn}]. Coulomb's law in the cgs system takes the form:
\begin{equation}F=\frac{Q_1Q_2}{r^2}\label{eq:coulomb}\end{equation}

\begin{figure}[bt]
\includegraphics[width=0.37\textwidth]{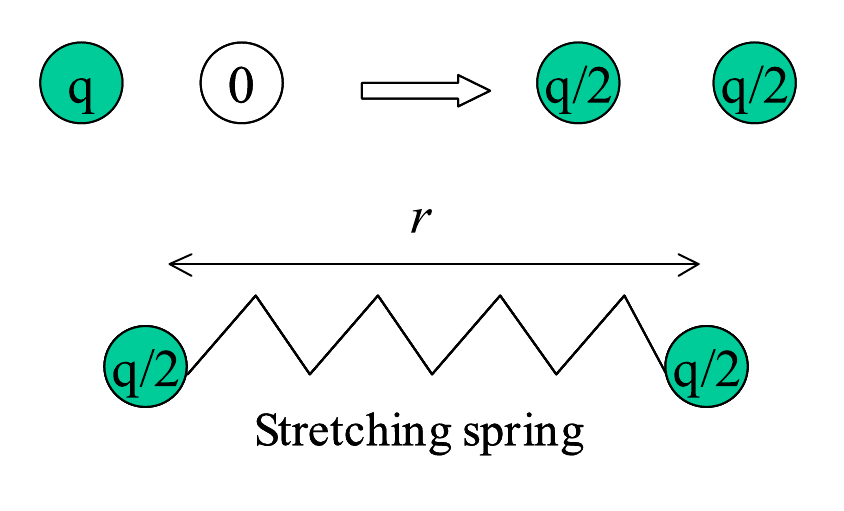}
\caption{A possible verification of Coulomb's law. \label{Fig:coulomb}}\end{figure}

The law does not only make a quantitative statement, but also introduces the dimensions of the electric charge in a most
natural way, based on the previously known dimensions of the mechanical quantities $F$ and $r$. The law can be mechanically verified by the following thought experiment (Fig. \ref{Fig:coulomb}): An initially charged (e.g. by rubbing) insulating sphere
is put in physical contact with another such sphere that is neutral. Since the two spheres are identical, the
charge will be redistributed equally between them (Fig. \ref{Fig:coulomb}, top). This consideration does not require any knowledge at all of
what charge is; it is based solely on symmetry. The result is that we have two identically charged spheres, whose
charges are $q/2$ regardless of what $q$ is.  As the spheres are attached to the ends of an elastic spring, they will repel each other,
thus stretching the spring (Fig. \ref{Fig:coulomb}, bottom). The string elongation can be translated into force $F$. We can use other
uncharged identical spheres, which by coming in contact with the previous charges will cause further charge fractionation:
$q/2\rightarrow q/4\rightarrow q/8..$, etc. (We note parenthetically that Coulomb used not an elastic but a torsion meter for measuring forces, but we would not like to get into these technical details.) Consequently, we can change the sphere with different charges without any knowledge of
what the physical origin of charge is. The aforementioned procedure provides a recipe for {\it mechanically} verifying the force, the charge, and the $r$
dependencies in the Coulomb law.

As written in
terms of dimensions, the Coulomb law yields: $$[Q]=[M]^{1/2}[L]^{3/2}[T]^{-1}$$
which connects the `mysterious' dimensions of charge to known mechanical dimensions. The latter representation enables dimensional estimates of all quantities that contain electrical charges.

As another example, let us consider Plank's constant: its dimensions can be based on the knowledge that its product with
frequency, $h\nu$, represents energy. The dimensions of the latter are easily derived from the kinetic energy $E=mv^2/2$, i.
e. $$[E]=[M][v]^2=[M][L]^2[T]^{-2}.$$ Taking into account $[\nu ]=[T]^{-1}$ yields $$[h]=[\hbar
]=[M][L]^2[T]^{-1}.$$

With the above in mind one can estimate the characteristic electron velocity $v_e$ in an
atom.  We start with the hydrogen atom and assume (hoping to verify later) that $v_e$ is well below the speed of light, making the latter irrelevant. The relevant quantities at hand remain $\hbar$, the electron charge $e$, and
the electron mass $m_e$ (the proton mass is irrelevant). Their general combination $\hbar ^{\delta}e^{\beta}m_e^{\gamma}$ must have the dimensions of
$[v]$, which yields $$[M]^{\delta}[L]^{2\delta}[T]^{-\delta}[M]^{\beta /2}[L]^{3\beta
/2}[T]^{-\beta}M^{\gamma}=[L]^1[T]^{-1}.$$  Setting equal the corresponding exponents on the left and right sides yields,
$$\delta +\beta /2+\gamma =0,\ 2\delta +3\beta /2=1,\ -\delta - \beta =-1.$$
From here, we get $\delta = -1$, $\beta =2$, $\gamma =0$, i. e. $$v_e\sim\frac{e^2}{\hbar}.$$

To estimate the importance of relativistic effects, we divide the latter characteristic velocity by the speed of light (in vacuum), which results in the dimensionless parameter $\alpha$ (fine structure parameter) given by:  \begin{equation}\alpha=\frac{e^2}{\hbar c}\approx
\frac{1}{137}.\label{eq:finestr}\end{equation} The fine structure parameter indeed describes relativistic effects in QM, which is the reason that it naturally appears in a number of specific problems.

Similarly, based on just dimensionality, the characteristic atomic radius can be estimated as:
\begin{equation}a_0\sim\frac{\hbar ^2}{m_e e^2}\label{eq:radiusa0}\end{equation} which can be used as an exercise for the students. Multiplying the latter  by the fine structure constant $\alpha$, we obtain the Compton length $L_c=\hbar/m_e c$, which equals the wavelength of a photon whose energy is
$m_e c^2$. The Compton length will be discussed again later in this manuscript.

For the calculation of the characteristic  energy of the electron, we can either follow the above procedure or use a shortcut: $$E\sim
m_e v_e ^2\sim\frac{m_ee^4}{\hbar ^2}.$$

It is standard in QM to represent system energies in terms of quantum numbers ($n$) that enumerate possible energy levels in ascending order. The above dimensional estimates alone do not show any dependency upon the quantum number $n$. However, it can be
readily restored by noting that $n$ always appears in combination with $\hbar$ according the De Broglie
quantization law (illustrated below in Sec. \ref{sec:DBQM}), and therefore can be accounted for by the substitution $\hbar\rightarrow n\hbar$. Also, for more complex
atoms with numbers $Z>1$, the $Z$ modification is achieved by noting that the attraction between the nucleus charge
$Ze$ and the electron is described by $Ze^2$, and therefore we can substitute $Ze^2$ instead of $e^2$ in the above equations (neglecting electron-electron interactions).

As another example of dimensional analysis we can consider the energy levels of a particle of mass $m$ in a quartic
potential well $U=Cx^4$, where $C$ is a constant. We can first find the dimensions of $C$, i. e.
$$[C]=[E]/[L]^4=[M][L]^{-2}[T]^{-2}.$$ and then consider a combination $\hbar ^{\delta}C^{\beta}m^{\gamma}$ that
can be written as
$$[M]^{\delta}[L]^{2\delta}[T]^{-\delta}[M]^{\beta}[L]^{-2\beta}[T]^{-2\beta}M^{\gamma}=[M][L]^2[T]^{-2}$$ where
the right hand side represents units of energy. Using the previous reasoning, equating dimensions and using $n\hbar$
instead of $\hbar$ leads to the result
$$E_n\sim n^{4/3}\left(\frac{\hbar ^4C}{m^2}\right)^{1/3}.$$

The energy spectrum for a particle in a magnetic field can also be derived from dimensional analysis. An interesting feature is that not all possible combinations of relevant parameters should be explored. Indeed, the Lorentz force on a particle of charge $e$ in magnetic field $B$, in the Gaussian system, is given by,
\begin{equation}\label{eq:lorentz}{\bf F}=\frac{e}{c}{\bf v}\times {\bf B},\quad {\rm i.e.} \quad F\sim \frac{e}{c}vB. \end{equation} We observe that the quantities $e$, $B$ and $c$ appear together as the combination $eB/c$.

In terms of dimensions,
$$ \frac{[e][B]}{[c]}=\frac{[F]}{[v]}=
\frac{[M][L][T]}{[T]^2[L]}=\frac{[M]}{[T]}.$$ Therefore
$$\Big[\frac{eB/c}{M}\Big]=\frac{1}{[T]}=[\omega].$$
Since $E = \hbar \omega$, we expect the energy to be of the form:
$E\sim \hbar (eB)/(cm) $.
According to our observations in the Dimensional Analysis section, we can restore the value of the principal quantum number $n$, such that \begin{equation}\label{eq:landau}E _n\sim n\hbar \frac{eB}{cm}. \end{equation}
That is the famous spectrum of Landau levels observed for electrons in a magnetic field.

\subsection{Using dimensions in differential equations}
In certain cases, dimensional arguments can help us analyze differential equations. We will start with the example of the Schr\"{o}dinger equation for a harmonic oscillator:
\begin{equation}-\frac{\hbar^2}{2m}\frac{d^2\Psi}{dx^2} + \frac{kx^2}{2}\Psi = E\Psi\label{eq:Shrond1}\end{equation}
We now introduce the dimensionless quantities $\tilde{x}$ and $\tilde{E}$ such that
$$ x= l\tilde{x} \quad , \quad E= \epsilon\tilde{E}$$
Eq. (\ref{eq:Shrond1}) will now become:
\begin{equation}-\frac{\hbar^2}{2m}\frac{1}{l^2}\frac{d^2\Psi}{d\tilde{x}^2} + \frac{kl^2\tilde{x}^2}{2}\Psi =\epsilon \tilde{E}\Psi\label{eq:Shrond2}\end{equation}
Since $l$ and $\epsilon$ remain yet undefined, one can set
\begin{equation}\label{eq:lE1}\frac{\hbar^2}{2m}\frac{1}{l^2}=\frac{kl^2}{2}=\epsilon, \end{equation}
after which all the quantities involved cancel out in Eq. (\ref{eq:Shrond2})
leading to a dimensionless form,
\begin{equation}
-\frac{d^2\Psi}{d\tilde{x}^2} + \tilde{x}^2\Psi = \tilde{E}\Psi .
\label{eq:Shrond3}\end{equation}
Since the latter does not have any parameters or numerical coefficients, it is natural to consider its independent variable in the range of $\tilde{x}\sim 1$, in which case  its solutions will have the form of eigenfunctions and eigenvalues of the order of unity, not of interest in qualitative analysis. The only relevant information is obtained from solving Eq. (\ref{eq:lE1}),
\begin{equation}\label{eq:harm}l^2=\sqrt{\frac{\hbar^2}{mk}} \quad, \quad \epsilon=\frac{k}{2}\sqrt{\frac{h^2}{mk}}\sim \hbar\omega,\end{equation}
where $\omega=\sqrt{k/m}$ is the classical frequency of a harmonic oscillator. Eq. (\ref{eq:harm}) presents known expressions for the localization length and quantum energy of a harmonic oscillator. Note that the above introduced replacement of  $\hbar\rightarrow n\hbar$ enables us to even obtain the equidistant spectrum $n\hbar\omega$ for a harmonic oscillator.

Similarly, the Schr\"{o}dinger equation for the hydrogen atom,
\begin{equation}-\frac{\hbar^2}{2m}\nabla ^2\Psi - \frac{e^2}{r}\Psi = E\Psi\label{eq:Shrond4}\end{equation} can be reduced to a dimensionless form.

Making the substitutions
$$ r= l\tilde{r} \quad , \quad E= \epsilon\tilde{E}$$
and equating:
\begin{equation}\label{eq:H}\frac{\hbar^2}{2m_e}\frac{1}{l^2}=\frac{e^2}{l}=\epsilon\end{equation}
yields the dimensionless equation:
\begin{equation}-\tilde{\nabla}^2\Psi - \frac{1}{\tilde{r}}\Psi = \tilde{E}\Psi,\label{eq:Shrond5}\end{equation}
Solving this equation will result in some numerical factors of the order of unity, which are of no interest here. In the meantime, solving Eq. (\ref{eq:H}) leads to the known parameters of the hydrogen atom,
\begin{equation} \label{eq:Hpar}l=a_0=\frac{\hbar ^2}{m_e e^2}, \quad E=\frac{m_e e^4}{\hbar ^2}.\end{equation}
On the same token, the angular momentum $\cal{L}$ is given by: $\cal{L}$$= m v_e l$. Additionally, using $n\hbar$ instead of $\hbar$ reproduces the hydrogen spectrum.

Equations other than Schr\"{o}dinger's  can be treated in a similar way. Consider, for example,
the heat flow (or diffusion) equation,
\begin{equation}\frac{\partial \eta}{\partial t} = \kappa \nabla^2\eta\label{eq:diffusion1}\end{equation}
where $\eta$ is a function of coordinates and time and $\kappa$ is the diffusivity.
Making the substitutions :
$$ r= R\tilde{r} \quad , \quad t= T\tilde{t}$$
we obtain the dimensionless equation:
\begin{equation}\frac{1}{T}\frac{\partial \eta}{\partial \tilde{t}} = \frac{\kappa}{R^2} \tilde{\nabla}^2\eta\label{eq:diffusion2}\end{equation}
where \begin{equation}R=\sqrt{\kappa T}\label{eq:diffusion3}\end{equation} The latter is the well-known relation between the diffusion distance and time.

It can be used as an exercise for the students to show that using dimensions with the wave equation, $\partial ^2f/\partial t^2=v^2\nabla ^2f$ leads to the relation $R=vt$ between the coordinate $R$ and time $t$.

We end this section by pointing at a somewhat different way of using dimensions with differential equations. This is a way that utilizes the dimensionality of differential operators (or integral operators when needed) in the same approximative manner that we already mentioned in section \ref{intro}. In particular, we can approximate:
\begin{equation}\frac{d}{dx}\sim\frac{1}{x},\frac{d^2}{dx^2}\sim\frac{1}{x^2},\nabla^2\sim\frac{1}{r^2}.\label{eq:approx}\end{equation}
In this case $x$ and $r$ should be understood as characteristic lengths. For example, equation (\ref{eq:Shrond1}) becomes:
$$-\frac{\hbar^2}{2m}\frac{1}{x^2} + \frac{kx^2}{2} = E.$$
Setting all the terms approximately equal to one another, we once more obtain Eq. (\ref{eq:harm}). (Note that equating the first two terms is the only way to define the characteristic length $x$ predicting a certain energy.)

Using the approximations: $\partial \eta / \partial t\sim \eta/t$ and $\nabla^2\eta\sim\eta/r^2$ in the diffusion Eq. (\ref{eq:diffusion1}) renders Eq. (\ref{eq:diffusion3}). Similar substitutions in the Schr\"{o}dinger equation for the hydrogen atom (eq. (\ref{eq:Shrond4})) yields relations (\ref{eq:Hpar}).

\section{De Broglie quantum mechanics}\label{sec:DBQM}
De Broglie postulated that every material particle can behave as a wave with its corresponding wavelength being
\begin{equation}
\lambda=\frac{h}{p}\label{eq:debroglie}\end{equation} where $h$ is Plank's constant and $p=mv$ is the particle's momentum. It has become apparent since then that his approach to QM is tantamount to the geometrical optics theory of light.

In reality, $\lambda$ as given by eq. (\ref{eq:debroglie}), is coordinate dependent because the momentum $p=\sqrt{2m(E-U)} $ includes the coordinate dependent potential energy $U$. However, in many cases, neglecting that coordinate dependence leads to reasonable approximations as shown next. Therefore, in what follows, we will approximate $U$ as a constant, effectively replacing $p$ with its average. Even in that simplified form, the De Broglie approach -often disdained in the standard physics curricula- remains as useful for practicing researchers as that of geometrical optics.

A more quantitative approach accounting for the wave nature of objects should use the wave equation that takes the form of the Schr\"{o}dinger equation in QM, and describes a wavefunction which is related to the probability density of the object.

\subsection{De Broglie quantization}\label{DeBroglie}
Treating a stationary state of a particle as a wave leads directly to the concept of standing particle-waves \cite{opentextbc}. The standing wave condition is that there must be an integral number of wavelengths set along any closed trajectory. We can use the aforementioned logic to derive a number of interesting conclusions.

\subsubsection{Rigid rotator}\label{sec:rr}
\begin{figure}[bt]
\includegraphics[width=0.32\textwidth]{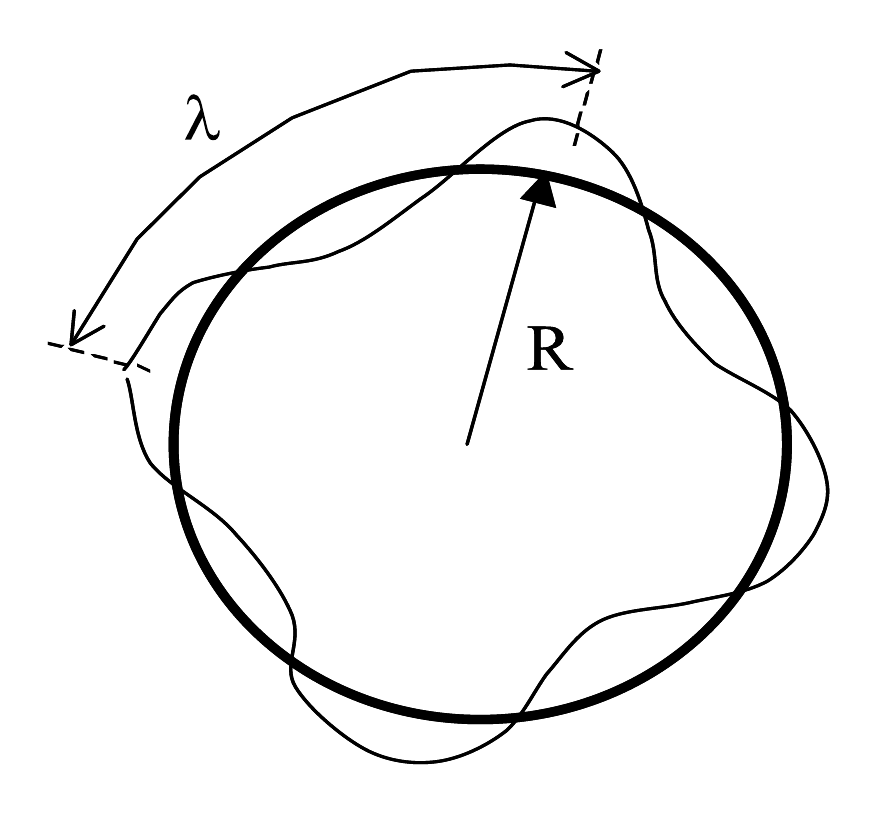}
\caption{A stationary wave of length $\lambda$ set along a closed circle trajectory for a particular case $n=4$.
\label{Fig:rotator}}\end{figure}

Let us consider the energy levels of a particle of mass $M$ that moves in a closed circular trajectory of radius $R$ (Fig. \ref{Fig:rotator}). Using the stationary wave condition
\begin{equation}\label{eq:rr}2\pi R=n h / p\end{equation}
where $n=1,2,3,..$ is an integer and the fact that its energy is of kinetic nature,
$E=Mv^2/2=p^2/2M$
yields
\begin{equation}
E=E_n\equiv n^2\frac{1}{8\pi ^2}\frac{h^2}{MR^2}\equiv n^2\frac{\hbar ^2}{2MR^2},\hfil
n=1,2,3...\label{eq:Erotator}\end{equation}
This result shows that the energy is quantized: $E_n=n^2E_1$, well known from standard quantum mechanical treatments \cite{zettili}.

As a practical implementation of the rigid rotator model we consider molecular rotations, in which case $R$ is of the
order of molecular length, i. e. the atomic (Bohr's) radius, $R\sim \hbar ^2/m_e e^2\sim 1$ \AA \quad where $m_e$ and $e$ are the
electron mass and charge respectively, and M is the atomic mass of the atom (which is approximately equal to the mass of the nucleus). Taking into account the expression for the characteristic atomic energy,
$E_a\sim m_e e^4/\hbar ^2\sim 10$ eV the result in Eq. (\ref{eq:Erotator}) can be presented in the form
$$E_n\sim n^2\frac{m_e}{M}E_a$$
with the famous adiabatic factor $m_e/M$ reflecting how the molecular dynamics are slower than the electronic ones. Assuming
$m_e/M\sim 10^{-4}$ (as a rough guide estimate) yields $E_1\sim 10^{-3}$ eV, which is the characteristic scale of
energy quantization in molecular rotations.

We note in passing that the characteristic energy is represented by the combination of the parameters $\hbar
^2/MR^2$, the only one having the dimension of energy, a result that can also be
established from dimensional analysis (cf. Sec. \ref{sec:dpq}).

\subsubsection{The hydrogen atom}
Referring again to Fig. \ref{Fig:rotator}, we assume that the electron in a hydrogen atom is bound by the Coulomb potential, e. g. its energy is
$$E=-\frac{e^2}{r}+\frac{p^2}{2m_e}$$
By adding the quantization condition of Eq. (\ref{eq:rr}), it is straightforward to derive the well known energy spectrum $E_n=-m_e e^4/2\hbar ^2n^2$, as well as the earlier noted results for the velocity and radius of its orbits, which can be also used as an exercise for the student.

\subsubsection{Particle in a one-dimensional potential box}\label{sec:QMBox}
\begin{figure}[bt]
\includegraphics[width=0.23\textwidth]{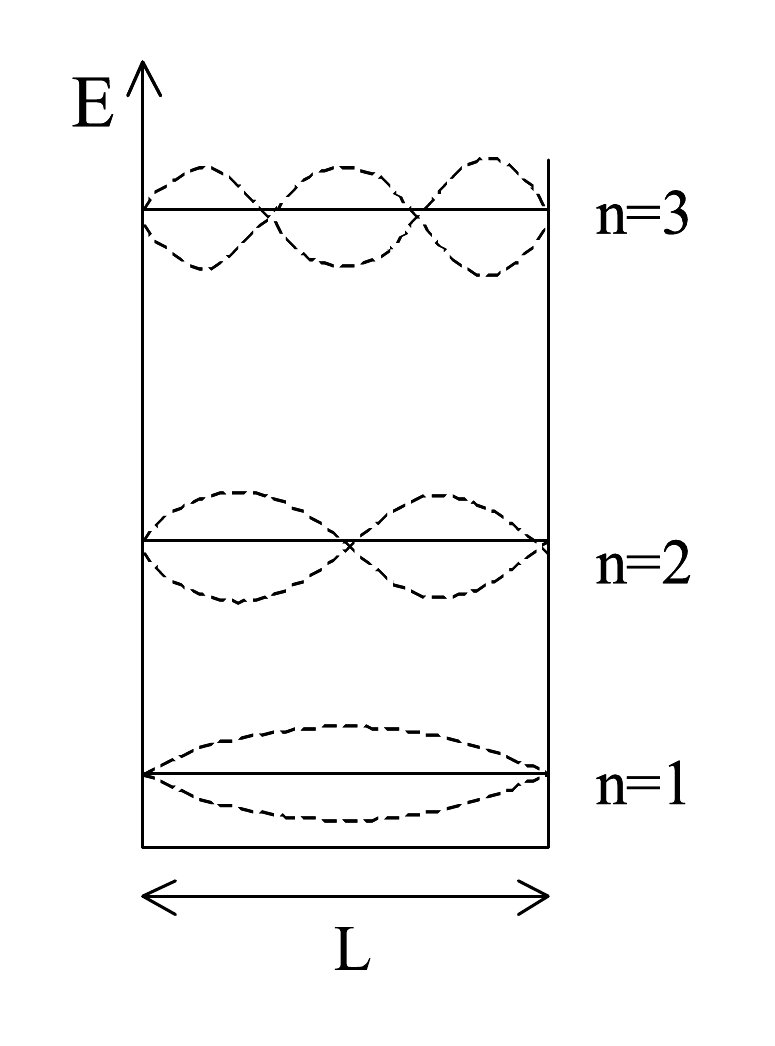}
\caption{Quantized energy spectrum and corresponding De Broglie waves in a potential box for
$n=1,2,3$.\label{Fig:box}}\end{figure}

Let us consider a particle of mass $m$ moving in one dimension between two walls separated by distance $L$. In this case, the closed trajectory corresponds to the particle moving back and forth in a length $2L$. Therefore the standing wave condition takes the form (Fig. \ref{Fig:box})
$$2L=n\frac{h}{p},\quad n=1,2,3....$$ The energy equation yields
\begin{equation}
E_n=\frac{p^2}{2m}=n^2\frac{\pi ^2}{8}\frac{\hbar ^2}{mL^2}.\label{eq:potbox}\end{equation} The same result follows by solving the Schrodinger equation. If the localization length $\ell$ is of atomic size, ($L\sim\hbar ^2/me^2 \sim 1$ \AA), then the
characteristic energy $E_1\sim  10$ eV is of the atomic energy scale as well.

For macroscopic motion, for example that of an electron localized in a metal cube of linear dimension $L=1$ cm, the
energy scale is lower by 16 orders of magnitude, $E_1\sim 10^{-15}$ eV, so that quantization becomes immaterial and can be neglected.

\subsubsection{Particle in a magnetic field}\label{sec:mf}
It is well known from classical mechanics that a particle of charge $e$ and mass $m$ moving at a speed $v$ in
a perpendicular magnetic field of induction $B$ will exercise circular motion with the {\it cyclotron} angular
velocity $\omega _c=eB/mc$ (in Gaussian units). Taking into account the De Broglie quantization, an integral
number of standing waves should be set along the trajectory length $2\pi R$. We can also use Newton's second law
for a particle under the Lorentz force and the kinetic energy equation:
\begin{equation}
\label{eq:Lor}n\frac{h}{mv}=2\pi r, \ \frac{e}{c}vB=m\frac{v^2}{R},\ E=\frac{mv^2}{2}.
\end{equation} It can be left as an exercise for the student to prove that these equations predict the equidistant discrete energy spectrum of Landau levels given in Eq. (\ref{eq:landau}) above.
%

\subsubsection{Harmonic oscillator}
\begin{figure}[bt]
\includegraphics[width=0.37\textwidth]{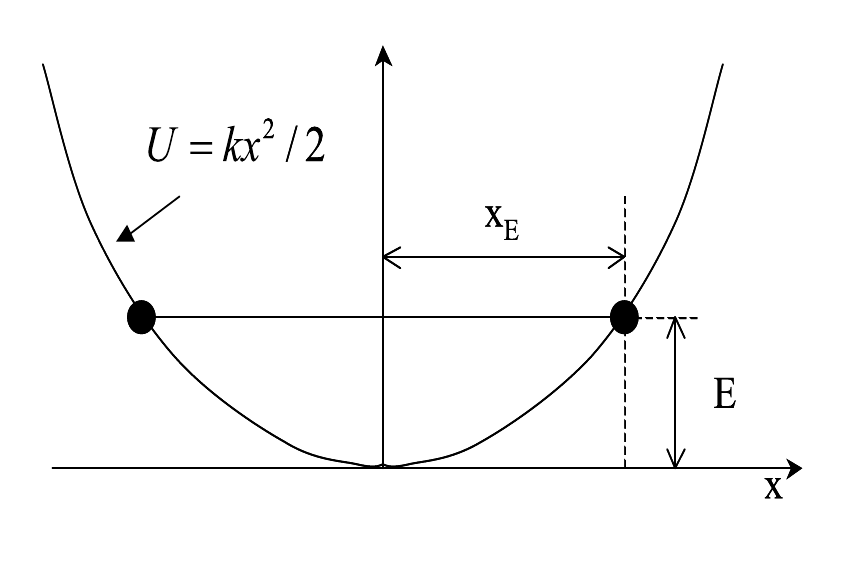}
\caption{Potential energy and turning points of a harmonic oscillator. \label{Fig:ho}}\end{figure}

We are now considering a mass $m$ attached to a spring of elasticity $k$ with potential energy $U(x)=kx^2/2$ (Fig. \ref{Fig:ho}). From classical mechanics, it exercises harmonic vibrations with frequency $$\omega =\sqrt{k/m}.$$ In its quantum description it also results in the energy quanta $\hbar\omega$, a result that will be derived next using the De Broglie quantization.

Considering the oscillator's turning points, its trajectory length is $L_{osc}=4x_E$ where $x_E=\sqrt{2E/k}$ as defined by the condition
$E=U(x)$. The quantization recipe takes the form: $$4\sqrt{\frac{2E}{k}}=n\frac{h}{mv}.$$

In this situation, the velocity is a function of the coordinate $x$ as given by the equation $v=\sqrt{2T/m}=\sqrt{2[E-U(x)]/m}$.  In the order of magnitude, we can however neglect this difficulty and simply eliminate $U(x)$ from the expression,
hence approximating $v=\sqrt{2E/m}$. Substituting the latter into the quantization condition yields:
\begin{equation}
E=n\frac{h}{8}\omega =n\frac{\pi }{4}\hbar \omega ,\label{eq:ho}\end{equation}
which is very close to $n\hbar\omega$.

Note that the correct answer takes considerably more mathematical effort (beyond the De Broglie approach) and returns the
energy quantum $E=\hbar \omega$ that is by $\approx 25$ \% higher than that in Eq. (\ref{eq:ho}), not
a very high price to pay for the incredible simplification provided by the above analysis. A related prediction of zero point
energy will be discussed in Sec. \ref{sec:zpv} below.

One important lesson learned from our harmonic oscillator example is that one can be adventurous enough to neglect the exact form of potential energy by approximating it as a constant (within certain limits). This approach will be further illustrated below (Sec. \ref{sec:uniff}).

As an interesting qualitative application of the harmonic oscillator, we can approximate  an atom as a uniformly charged sphere where the negatively charged electron oscillates under the influence of the Coulomb force (the `gel' model). The average positive charge density $\rho$ is given by $\rho \sim e/a_0^3$, where $a_0$ is the atomic radius.  The corresponding potential energy at distance $r$ from the center is estimated as $\rho er^2$. Expressing the latter in the form $kr^2$ gives: $$k=\frac{e^2}{a_0^3}.$$ If we substitute our result in the formula for the harmonic oscillator energy, we obtain
$$\hbar\omega =\sqrt{\frac{k}{m_e}}=\sqrt{\frac{\hbar ^2}{m_ea_0^2}\frac{e^2}{a_0}},$$
which is the geometrical average of the electron kinetic energy $\hbar ^2/m_e a_0^2$ [cf. Eq. (\ref{eq:potbox})] and the Coulomb
potential energy $e^2/a_0$. At this point we can introduce a useful example of ``guessing": whenever we deal with
unknown kinetic and potential energies, it is reasonable to assume that they are comparable in absolute value (we can check that for the aforementioned examples of the hydrogen atom or harmonic oscillator). Applying the latter `rule of thumb' gives $\hbar ^2/m_e a_0^2\sim e^2/a_0$
which allows us to derive,
 $$\hbar\omega\sim \frac{m_e e^4}{\hbar ^2},$$ in agreement with previous equations.

\begin{figure}[bt]
\includegraphics[width=0.42\textwidth]{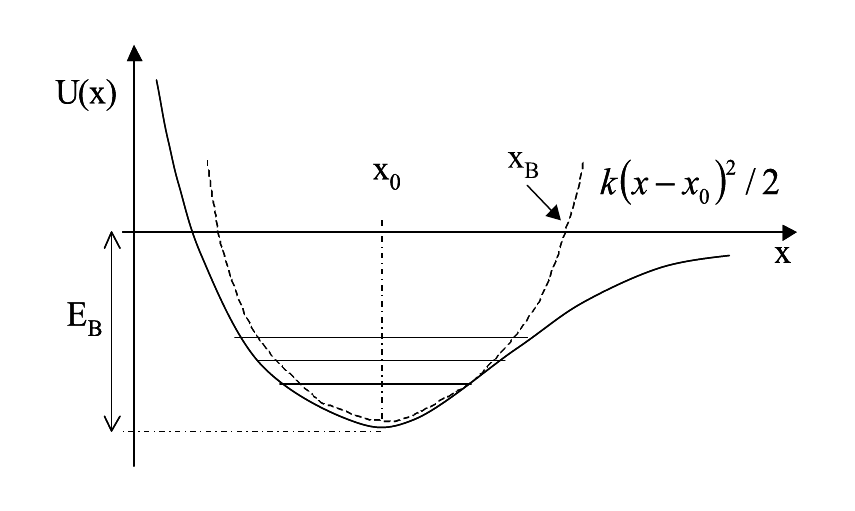}
\caption{Potential energy and its harmonic approximation (dashed) for a two-atomic molecule.
\label{Fig:mol}}\end{figure}

We will consider next another useful application of the harmonic oscillator: molecular vibrations. Shown in Fig. \ref{Fig:mol} is the familiar general
shape of the potential energy in a 2-atomic molecule having an equilibrium interatomic distance $x_0$.
 The molecular binding energy $E_B$ is of the order of the characteristic energy of
the electrons moving back and forth between the two atoms and binding them together. In a very rough
approximation we can estimate $E_B\sim \hbar ^2/m_ex_0^2$.  The spring constant describing the molecule in the
harmonic approximation (parabola in Fig. \ref{Fig:mol}) can be estimated from the observation that the molecule
has to be stretched considerably to overcome the bonding energy and break into individual atoms. This means
that in the harmonic approximation the breaking point $x_B$ can be defined such that $k(x_B-x_0)^2/2=E_B$ where $x_B-x_0\sim
x_0$, which gives $k\sim E_B/x_0^2$. The vibrational quantum energy corresponding to the {\it atomic} mass $M$
is estimated as:
\begin{equation}
\hbar\omega \sim\hbar\sqrt{\frac{E_B}{Mx_0^2}}=E_B\sqrt{\frac{m}{M}}.\label{eq:molvib}\end{equation}

\begin{figure}[bt]
\includegraphics[width=0.35\textwidth]{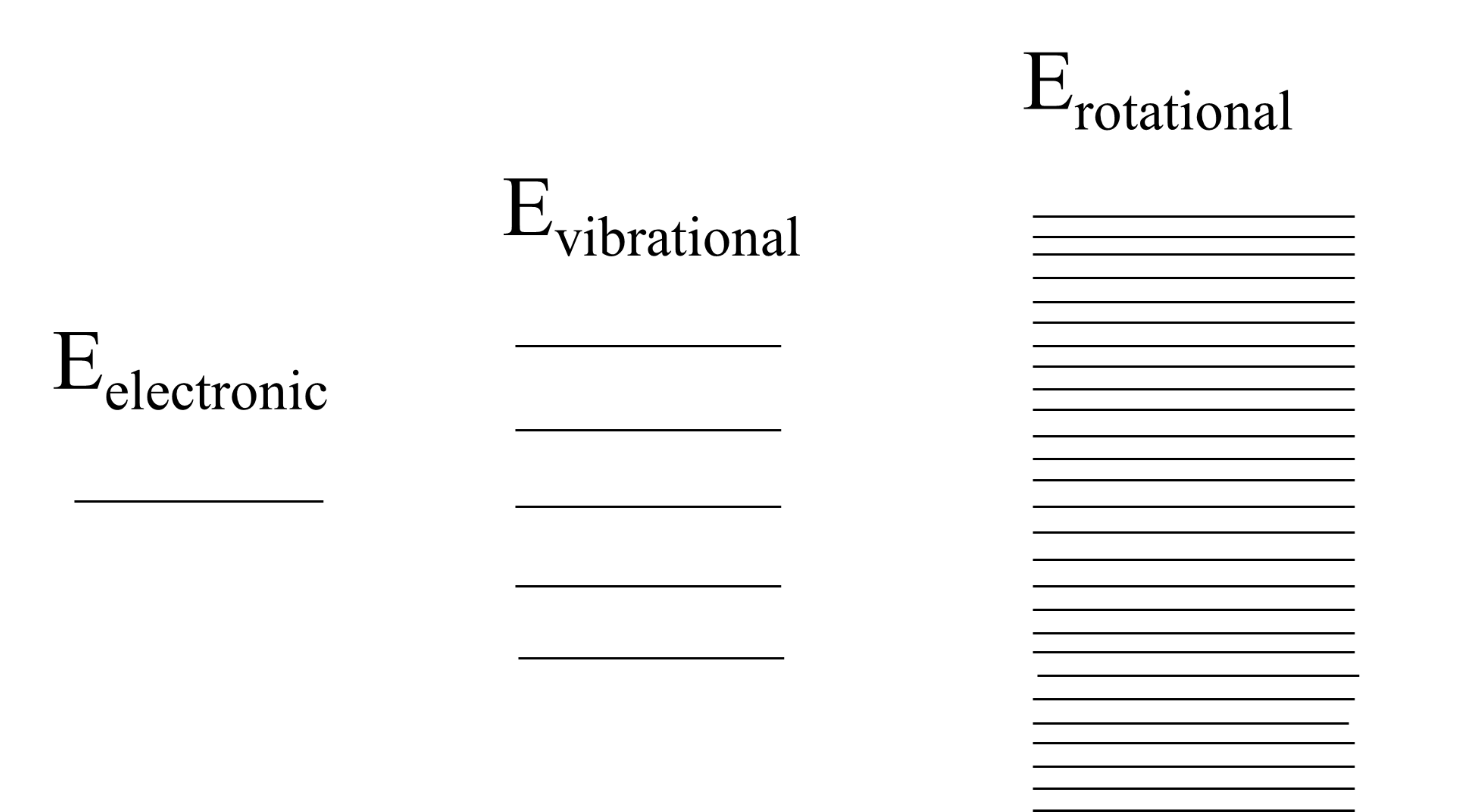}
\caption{The complex nature of molecular energy spectra containing the hierarchy of electron, vibrational and
rotational spectra (not to scale). \label{Fig:molspec}}\end{figure}

We note again the presence of the ubiquitous adiabatic factor $m_e/M\ll 1$, but this time under the square root, as opposed to the
case of molecular rotations. As a result, we arrive at the following hierarchy between the
characteristic energies of molecular rotations, vibrations, and the electron energy $E_{el}\sim E_B$:
\begin{equation}
E_{rotation}\sim\sqrt{\frac{m_e}{M}}E_{vibration}\sim\frac{m_e}{M}E_{el},\label{eq:hier}\end{equation} which can be also proven rigorously. The result illustrates the rather complex structure of the universally observed molecular energy spectra
graphically sketched in Fig. \ref{Fig:molspec}.

\subsubsection{Quantization in a triangular well}\label{sec:uniff}

\begin{figure}[bt]
\includegraphics[width=0.25\textwidth]{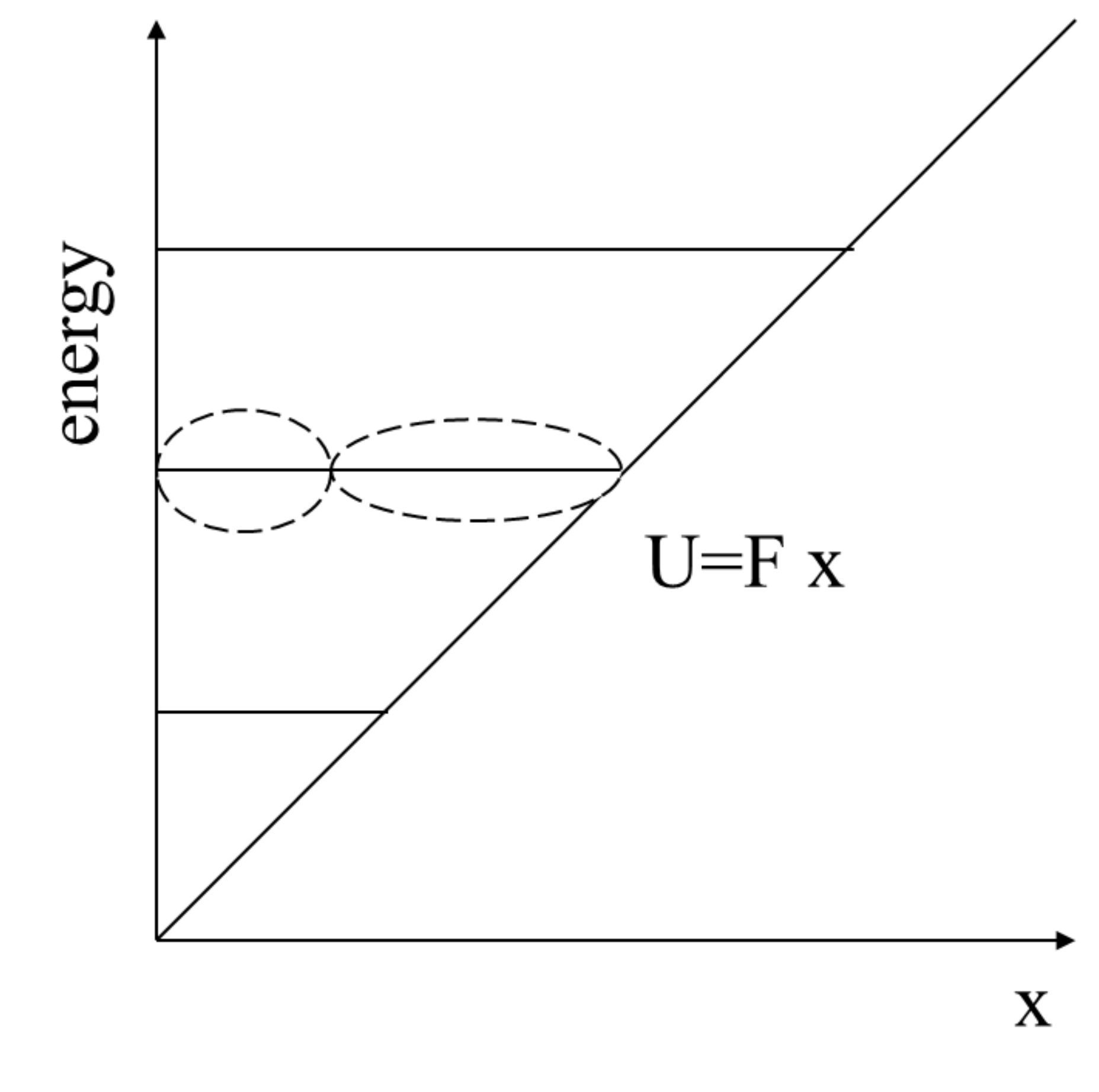}
\caption{The potential energy of electrons attracted to an impenetrable surface (vertical line). Horizontal lines represent several energy levels of electrons, the dashed lines represent a standing De Broglie wave. \label{Fig:uniff}}\end{figure}

This example is important because of its multiple practical applications in modern electronic devices, where the electrons are attracted to an impenetrable surface by a uniform electric field as illustrated in Fig. \ref{Fig:uniff}.
Using the same reasoning as in the previous example, we find the turning points $x=0$ and $x=E/F$ by assuming that the potential energy is of the form $U=Fx$. The trajectory length becomes $2E/F$. We then use the De Broglie quantization condition and approximate $U=0$ when $0<x<E/F$, i. e.
\begin{equation}\frac{nh}{mv}=2\frac{E}{F}, \quad E=\frac{mv^2}{2}.\end{equation}
Solving the latter for energy and neglecting numerical multipliers yields the spectrum
\begin{equation}\label{eq:uniff}E_n\sim\left(\frac{F^2\hbar ^2n^2}{m}\right)^{1/3}.\end{equation}

A more exact analysis \cite{zeghbroeck} ends up with Airy functions that provide analytical solutions in the case of large enough quantum numbers $(n\gg 1)$ and can be obtained from our Eq. (\ref{eq:uniff}) by replacing
$$n^{2/3}\rightarrow \left[(3\pi /4)(n-1/4)\right]^{2/3},$$
which is indeed an insignificant correction.

As a straightforward exercise, the result in Eq. (\ref{eq:uniff}) can be obtained from dimensional analysis along the lines of Sec. \ref{sec:DE}.

\subsection{De Broglie tunneling}

\begin{figure}[bt]
\includegraphics[width=0.50\textwidth]{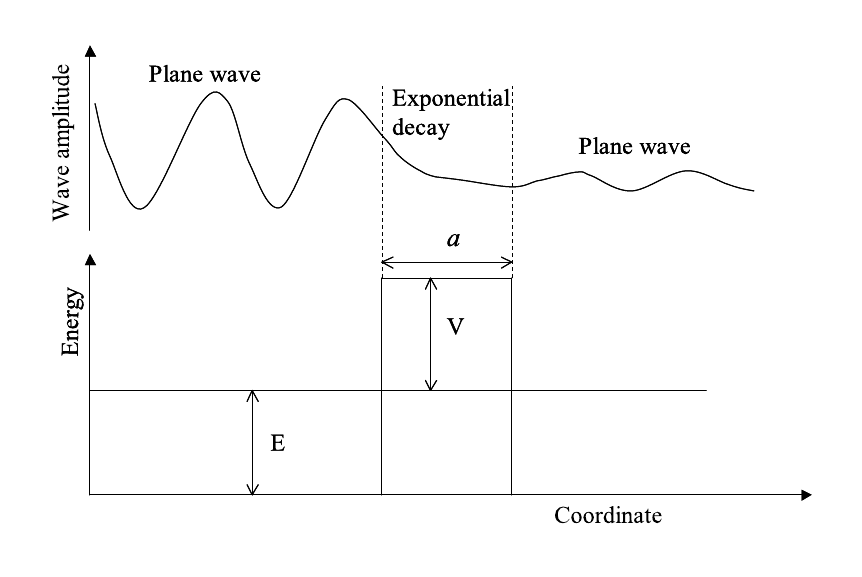}
\caption{Tunneling of a particle with energy $E$ incident on a potential barrier of height $V+E$. The barrier
region is classically forbidden.\label{Fig:tunn}}\end{figure}

 For a particle of mass m incident on a potential barrier (as illustrated in Fig. \ref{Fig:tunn}), the De Broglie wave function is:
\begin{equation}\psi =C\exp(ikx),\quad C=const\label{eq:dbw}\end{equation} in the classically allowed regions to the left and to the right of
the barrier. The wavefunction represents a plane wave with the wavenumber

$$k=i\sqrt{\frac{2m(U-E)}{\hbar ^2}}=i\sqrt{\frac{2mV}{\hbar ^2}}$$
where $i=\sqrt{-1}$ and $V$ is the barrier `cap' above the particle energy. Substituting the latter in the
expression for the De Broglie wave in Eq. (\ref{eq:dbw}) results in the exponentially decaying function
$$\psi =C\exp\left(-\sqrt{\frac{2mV}{\hbar ^2}}x\right)\equiv C\exp\left(-\frac{x}{\lambda}\right),
\quad \lambda=\frac{\hbar}{\sqrt{2mV}}.$$ This decay is illustrated in Fig. \ref{Fig:tunn}; note that the characteristic
decay length $\lambda$ is a direct relative to the De Broglie wavelength corresponding to the imaginary momentum of a particle with kinetic energy $V$.

The above exponential decay results in an exponentially small amplitude for the transmitted wave. The corresponding wavefunctions are:

$\psi =C_1\exp(ikx)$ for the incident wave,

$\psi =C_2\exp(ikx)$ for the transmitted wave,

and

$$\frac{C_2}{C_1}=\exp\left(-\frac{a}{\lambda}\right)$$ with $a$ being the barrier width.

Considering the analogy to EM waves, the observed quantity is characterized not as the wave's amplitude, but rather
as the square of the amplitude. Correspondingly, the square of the above ratio will describe the probability of tunneling,
\begin{equation}P=\exp\left(-\frac{2a}{\lambda}\right)=\exp\left(-2a\sqrt{\frac{2mV}{\hbar
^2}}\right).\label{eq:tunn}\end{equation}

Two comments are in order here: First, instead of the EM wave analogy, one could refer to the standard interpretation of QM where the square of the absolute value of the wave function describes the probability. Secondly, the above considerations neglect certain features analyzed in regular QM, such as barrier reflection and above the barrier transmission.

\subsubsection{Coin tunneling through a wall}
The following example will deal with intentionally classical objects. Let us consider a $\sim 1$ g mass dime trying to tunnel through a $\sim
1$ cm thick wall. We shall see that the probability of such tunneling is so small that it can be altogether neglected.

In our example, the barrier length is given as $a\sim 1$ cm, however, we need to yet estimate the tunneling length. We will utilize the interpretation that the barrier will have to  do work in order to increase the particle's energy
above the barrier, which is tantamount to the work of increasing the spacing of the wall atoms enough, to allow the
dime in. Such a process requires breaking the molecular bonds so as to make a necessary opening of volume
$\Omega \sim 1$ cm$^3$. Breaking $\sim 10^{14}$ molecular bonds in the surface enveloping such a volume requires an
energy of $\sim 10^{14}$ eV, assuming a binding energy of $\sim 1$ eV per molecule. Hence, $V\sim 10^{14}{\rm eV}\sim 100$
erg. Substituting this into the expression for the tunneling length gives $$\lambda\sim\frac{10^{-27}}{\sqrt{100}}\sim
10^{-28}\quad {\rm cm}.$$ The corresponding tunneling probability, $P\sim\exp(-10^{28})$ is small beyond our
imagination and can be neglected.

\subsubsection{Electron tunneling in diatomic molecules}

\begin{figure}[bt]
\includegraphics[width=0.50\textwidth]{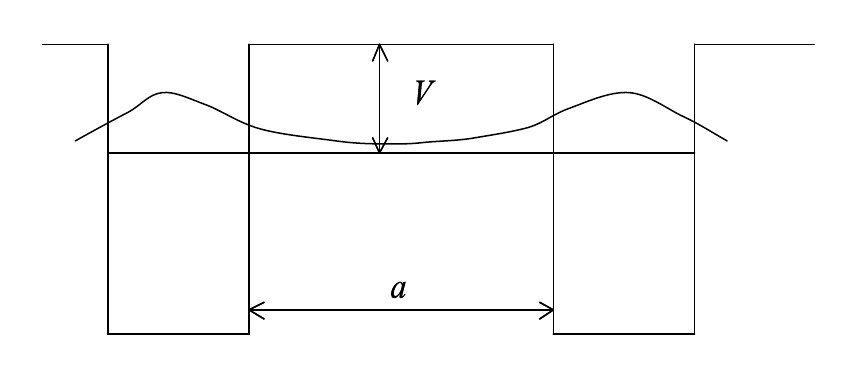}
\caption{Double-well potential representing atomic potentials in diatomic molecules and an electron (ground) wave function
equally distributed between them.\label{Fig:dam}}\end{figure}

A double-potential well (Fig. \ref{Fig:dam}) can be used to represent atoms in diatomic molecules that are separated by a
potential barrier which an electron can overcome by tunneling. In this case, the typical parameters are of the atomic
scale, $a\sim 2-3$ \AA \quad and $V\sim 1$ eV. The corresponding tunneling length,
$$\hbar /\sqrt{m_e V}\sim 10^{-27}/3\sqrt{10^{-27}10^{-12}}\sim 3\cdot 10^{-8}\quad {\rm cm}$$ is comparable to
the molecular size. Hence, the probability of tunneling $P\sim \exp(-2a/l)$ is not very small, i.e. $P\sim
0.3-0.1$, and the electron moves between the two atoms quite frequently thereby binding them into a molecule.

\subsection{The uncertainty principle}
\begin{figure}[bt]
\includegraphics[width=0.37\textwidth]{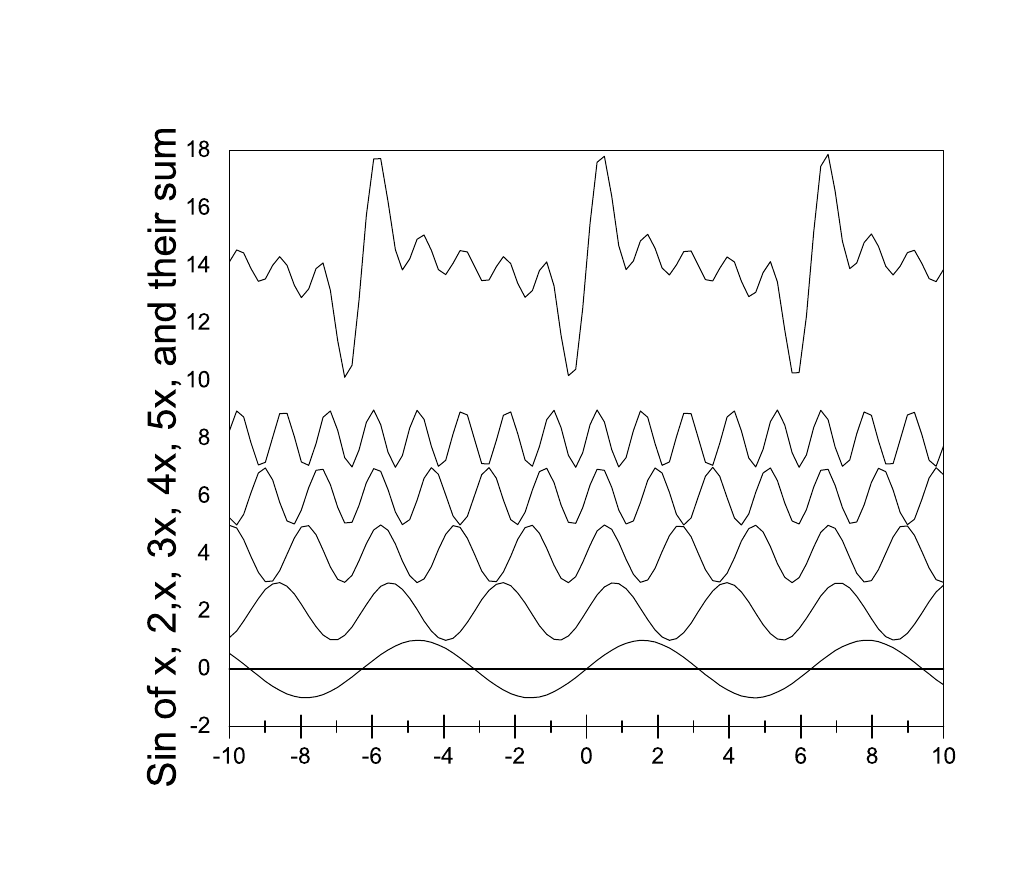}
\caption{Wavelets obtained by superposition of five trigonometric functions with the same weights. \label{Fig:wavelet}}\end{figure}

De Broglie waves can be used to describe a localized entity (wavelet) that will represent a particle at a more
intuitive level. This can be achieved by adding many plane waves, and results in amplifications and
cancellations in space as illustrated in Fig. \ref{Fig:wavelet}.
While this is  mathematically treated through Fourier analysis, we are hereby proposing a more intuitive approach \cite{wichmann}.

Let us consider a sum of cosine functions, where the wave number changes from $k$ to $k+\Delta k$ such that, $$\int _k^{k+\Delta
k}\cos (kx)\frac{dk}{\Delta k}=\frac{\sin(x\Delta k/2)}{x\Delta k/2}\cos[(k+\Delta k/2)x].$$ We observe that all
the participating plane waves cancel each other when $x\Delta k=2\pi$. The next point of cancelation, $x+\Delta x$ happens when $(x+\Delta x)\Delta k=4\pi$. From Fig. \ref{Fig:wavelet} we can define the width
$\Delta x$ of the wavelet:
\begin{eqnarray}4\pi &=& (k+\Delta k)(x+\Delta x)-k(x+\Delta x)= x\Delta k+\Delta x\Delta k \nonumber \\ &=& 2\pi +\Delta x\Delta k\nonumber \end{eqnarray} therefore,
$$\Delta x\Delta k=2\pi.$$

We can interpret the above point of mutual cancellations as the wavelet boundaries or the boundaries of the minimum
interval of particle localization. Considering several wavelets would increase the
particle dimensions, therefore we can replace the $=$ sign in the latter equation with $\gtrsim$.
Finally, using the De Broglie wavelength equation and ignoring the factor of $2\pi$, we can now rewrite the result as
\begin{equation}\Delta x\Delta p\gtrsim \hbar .\label{eq:uncert}\end{equation}
In the following applications we will use its `truncated' version, $\Delta
x\Delta p\sim \hbar$.

\subsubsection{Kinetic energy of a localized particle}

Consider a particle localized in a region of linear size $l$, which we now interpret as the uncertainty in
the particle's coordinate. Correspondingly, the uncertainty in its momentum is $\Delta p\sim \hbar /l$. Since the
momentum cannot be less than its uncertainty, we use the latter as the momentum estimate, $p\sim \hbar /l$. The
kinetic energy becomes
\begin{equation}\label{eq:kine}E=\frac{p^2}{2m}\sim \frac{\hbar ^2}{ml^2},\end{equation} consistent with the earlier result in Eq. (\ref{eq:potbox}). That is fully consistent with our previous estimates in Sec. \ref{sec:dpq} and \ref{sec:QMBox}.


Our result can be proven useful in many applications. For example, if we are given the
characteristic nuclear radius of $l\sim 10^{-13}$ cm and the binding energy of $U\sim 10$ MeV, is it possible for an
electron to be part of a nucleus? The answer is `No', because the kinetic energy of such a strongly localized
electron, $$E\sim\hbar ^2/ml^2\sim 10^{-54}/10^{-27}10^{-26}{\rm erg}\sim 10^{11}{\rm eV}$$
greatly exceeds the depth $U\sim 10^7$ eV of the available potential well (estimated as a typical binding energy in a nucleus). Hence, the latter cannot accommodate an electron: electrons cannot be a part of atomic nuclei.

\subsubsection{Zero point vibrations}\label{sec:zpv}
In classical mechanics, a particle at rest is always localized at the point of minimum potential energy, which  -in the case of the harmonic oscillator: $U=kx^2/2$- is at the point $x=0$. However, in QM, the uncertainty principle makes such localization impossible, giving rise to a minimum energy $E_{min}$.

Let us consider the sum of the potential and kinetic energies of a harmonic oscillator,
$$E(x)=U+K=\frac{kx^2}{2}+\frac{\hbar ^2}{2mx^2}$$ (where we keep the factor of 2 in the kinetic energy to make the
terms arithmetically similar). The kinetic term ($\propto 1/x^2$) -based on eq. (\ref{eq:kine})- does not let the particle decrease its
potential energy to the minimum value of 0 at $x=0$. Furthermore, there is an optimum value
$$x_0=\left(\frac{\hbar ^2}{mk}\right)^{1/4}=\frac{\hbar}{\sqrt{m\hbar\omega}}$$ obtained by the
minimization of $E(x)$ through $dE/dx=0$ and compromising between the opposite trends of the kinetic and potential
energy terms; note that $x_0$ is now represented by the familiar De Broglie wavelength. Substituting $x_0$ back
into $E(x)$ gives the minimum energy $$E_{min}=\hbar\omega .$$ This minimum energy estimated to within the
accuracy of a numerical coefficient corresponds to the $n=0$ term in the exact expression $E=\hbar \omega (n+1/2)$
for the energy of a harmonic oscillator.

\subsubsection{Diffraction of particles}

\begin{figure}[bt]
\includegraphics[width=0.28\textwidth]{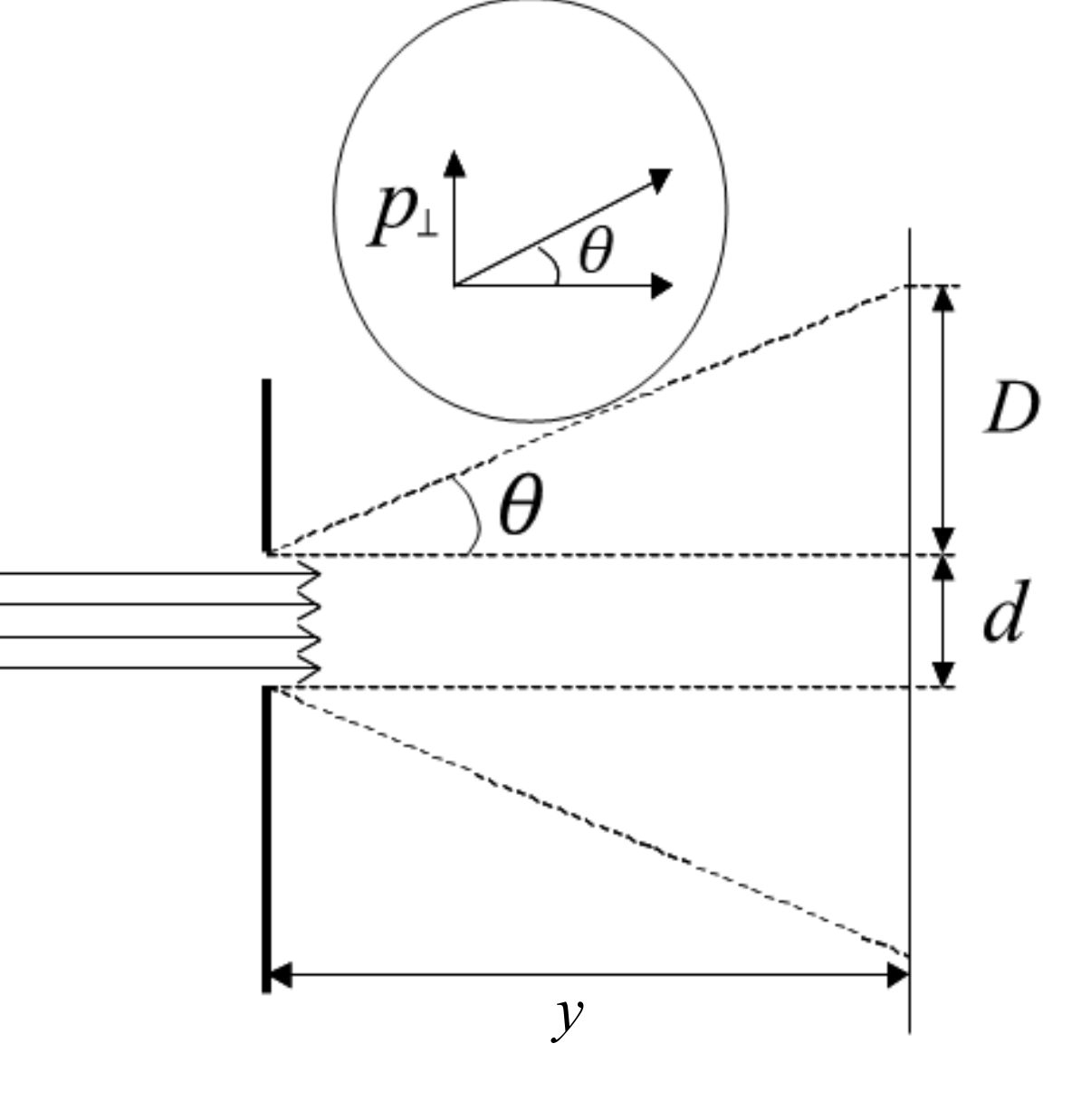}
\caption{Diffraction of particles passing through a slit where $d$ is the geometrical size of the slit and $D$ is the
characteristic diffraction pattern size. Shown in the inset is the vector of the particle's momentum illustrating its
perpendicular component. \label{Fig:diffr}}\end{figure}

Let us consider a  beam of particles \cite{wichmann} of momentum $p$ that passes through a slit of diameter $d$ and is registered by a screen at
distance $y$ (Fig. \ref{Fig:diffr}).
While passing through the slit, each particle's coordinate is defined to the accuracy of $d$, which, according to the
uncertainty principle, predicts the uncertainty in its perpendicular momentum component $\Delta p_{\bot}\sim \hbar
/d$. Considering that the momentum cannot be smaller than its uncertainty, the latter will also serve as an estimate for
the particle's perpendicular momentum, therefore the angle $\theta$ characterizing the particle's deflection can be
estimated as $$\theta\sim \frac{p_{\bot}}{p}\sim \frac{\hbar}{pd}\sim\frac{\lambda}{d}$$ where $\lambda$ is the
particle's De Broglie wavelength. Note that this result is remarkably similar to the wave theory's classical result for the
diffraction angle of light from a slit.

From the latter expression, the diffraction pattern size is estimated as $D\sim \theta y$, so that the total size
of the light spot is $$d+2D\sim d+y\frac{\lambda}{d}.$$ The expression has a minimum at $d=\sqrt{y\lambda}$. For the case of
smaller $d$ the diffraction makes the spot larger, while for larger $d$, it is the geometrical image that makes the spot larger.

\subsubsection{Quantum conductance}\label{sec:cq}
Ohm's law holds for as long as the system dimension $a$ significantly exceeds the electron mean free path. In the opposite limiting case of small systems, the electrons do not lose their respective energies by collisions. Instead, a certain resistance remains measurable in the form of a quantum conductance. The nature of it can be qualitatively related to the uncertainty principle, which allows enough uncertainty in the energy, so as to allow its measurable dissipation.

We now recall that conductance is a quantity reciprocal of the electric resistance. Quantized conductance occurs in ballistic conductors where the electron mean free path is much larger than the length of the system, so the electrons traverse it without collisions. Here we point at a qualitative derivation of quantized conductance conceptually close to that in Wikipedia \cite{condquan}.

The electric current due to one electron is presented as $I=e/t$ where $t\sim a/v$ is the electron travel time, $v$ is the electron velocity, and $a$ is the system length. Corresponding to the latter is the momentum uncertainty $\Delta p\sim \hbar /a$, and the energy uncertainty  $\Delta E= \Delta (p^2/2m)=v\Delta p$. Combining  these equations yields $I\sim q\Delta E/\hbar$. We now allow the electron to change its energy by $\Delta E$ and interpret that change as related to the electric potential difference across the system, $\Delta E=eV$. As a result, $I=(e^2/\hbar )V$ takes the form of the Ohm's law with the quantum conductance $e^2/\hbar$. The latter is close to the correct result $2e^2/h$.

\section{Other qualitative estimates in quantum mechanics}\label{sec:other}

\subsection{Spin-orbit interaction}\label{sec:soi}
\begin{figure}[bt]
\includegraphics[width=0.25\textwidth]{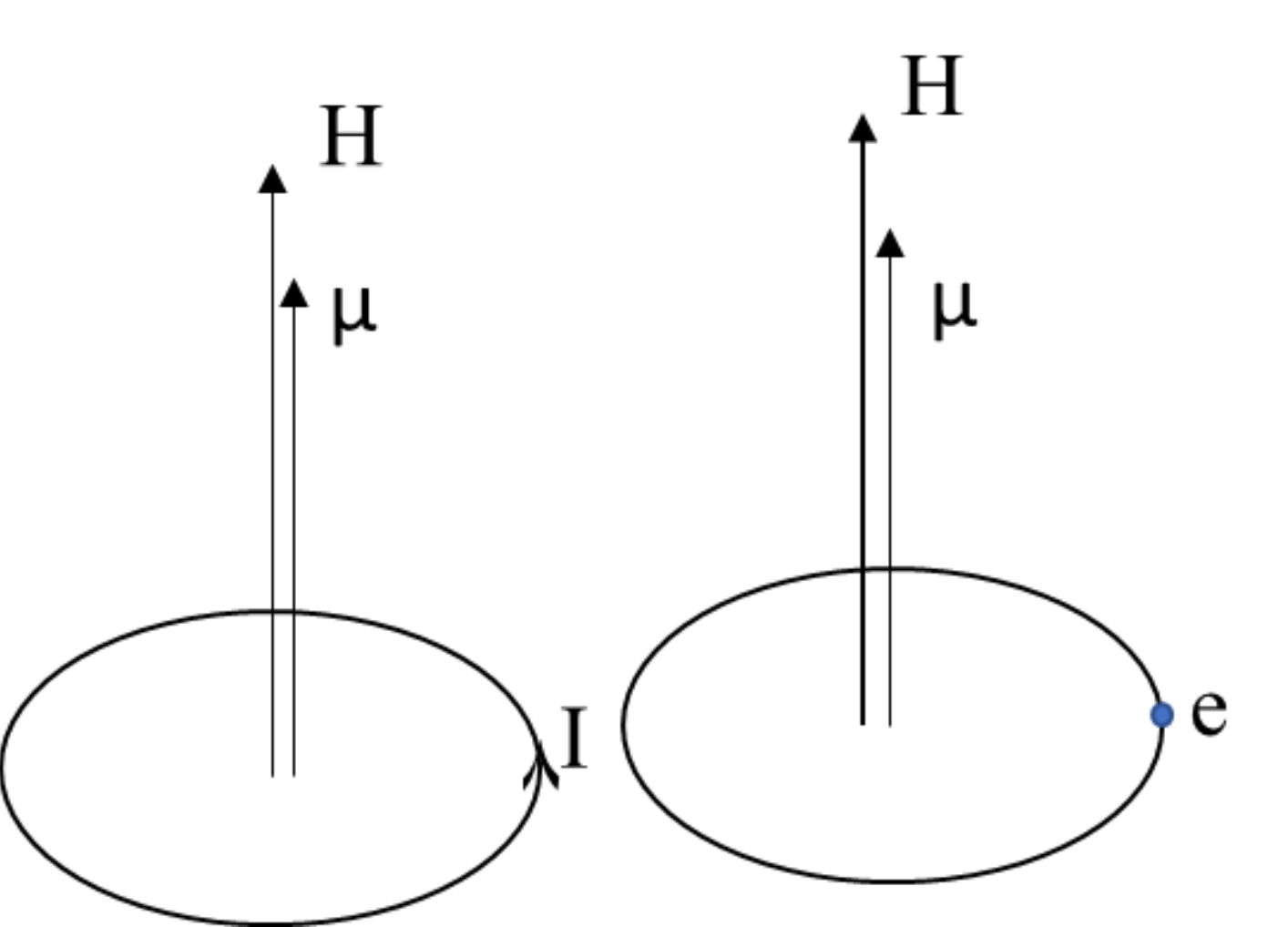}
\caption{Magnetic field and magnetic moment of a coil carrying current I. Magnetic field and magnetic moment of an electron e.
\label{Fig:mmoment}}\end{figure}

Classical electrodynamics tells us that moving particles generate a magnetic field, therefore, our electron can be also
seen as a course of a magnetic field. (We can remind ourselves of a circular coil which carries current; the current generates
a magnetic field in the direction shown in Fig. \ref{Fig:mmoment}).

 We also know that a circular coil can be assigned a magnetic moment $\mu \sim (A I)/c$, where $A$ is the area of the
coil, $I$ the current that it carries and $c$ is the speed of light.
In the same manner, the electron can also be assigned a magnetic moment $\mu$. The area $A$ is now: $A \sim r^2$ and the current $I = e/T$ , where $e$ and $T$ are the electron's charge and period.  Therefore:
\begin{equation}
\mu \sim \frac{r^2 e}{c T}.\label{eq:dmoment}\end{equation}
Using $v = r/T$ and the definition of angular momentum $\cal{L}$$= m v r$, we have: $$\mu \sim \frac{e}{m_ec}\cal{L} .$$ The combination $e/mc$ is called the gyromagnetic ratio.
The magnetic moment that we just discussed was due to the orbital motion of the electron. We can also assign an intrinsic spin angular momentum $\mu_s \sim (e / m_ec) s$, where s in the spin angular momentum.
Knowing that the magnitude of spin can be $\pm \hbar / 2$, we now have:
\begin{equation}
\mu_s \sim \frac{e\hbar}{m_ec}\label{eq:dmoment1}\end{equation}
(The quantity $e\hbar/mc$ is known as the Bohr magneton).

Let us now calculate the magnetic field that this electron produces. We can write down the equation for the radius of an electron around a nucleus that has $Z$ protons by using equation (\ref{eq:radiusa0}) and substituting $Ze$ instead of $Z$, and $n\hbar$ instead of $\hbar$ as shown in section \ref{sec:dpq}:
$$r \sim \frac {n^2\hbar^2}{m_eZe^2}.$$
We also know that the magnetic field of a coil with radius $r$ that carries current $I$, at the center of the loop is given by: $H \sim   I /cr$. In the case of the electron $ I = e/T$ and $ T = r / v$. From here we find:
\begin{equation}
H \sim \frac{e^7 Z^3 m_e^2}{c n^5 \hbar^5}.\label{eq:efield}\end{equation}

The product of $\mu$ and $H$ from Eqs. (\ref{eq:dmoment1}) and (\ref{eq:efield})  multiplied with the additional factor of Z to account for all the electrons of the atom,  yields the spin-orbit energy:
\begin{equation}
H \cdot \mu_s \sim \frac{Z^4 m_e e^8}{n^5 \hbar^4 c^2}\label{eq:spinorbit}\end{equation}

Our latter result coincides with that given in many standard textbooks. For example Zettili \cite{zettili} derives for the case of hydrogen atom (Z=1) the following spin-orbit energy:
\begin{equation}
\hat{H}_{so}= \frac{e^2}{2m_e^2c^2}\frac{1}{r^3} \overrightarrow{S} \cdot \overrightarrow{\cal{L}}. \label{eq:Zettili}\end{equation}
Using Eq. (\ref{eq:radiusa0}) for the radius of the electron (with $n\hbar$ instead of $\hbar$), $s \sim \hbar$ and $\cal{L}$$\sim n \hbar$ gives us
\begin{equation}
\hat{H}_{so} \sim \frac{m e^8}{n^5 c^2 \hbar^4}\label{eq:zettili1}\end{equation}
which is consistent with Eq. (\ref{eq:spinorbit}).
We also notice that Eq. (\ref{eq:spinorbit}) shows a $Z^4$ dependence for the spin-orbit interaction energy. This result is valid for a hydrogenic atom without considering screening effects and has been experimentally verified for low Z atoms. For higher Z atoms, screening effects have to be considered  \cite{shavanas,landauqm} and the spin-orbit interaction energy is proportional to $Z^2$.

\subsection{Transition rates}\label{sec:trara}
 Spontaneous emission occurs when -in the absence of radiation- an atom transitions to a lower state. We will use dimensional analysis to calculate the power emitted during spontaneous transitions, in a similar manner to that in reference [20]. Since power is defined as energy per unit time, its dimensions are:$$[P]=[E]/[T]=[e^2/L]/[T]=[e^2]/[L][T]$$ where e is the charge of an electron.

\begin{figure}[h!]
\includegraphics[width=0.20\textwidth]{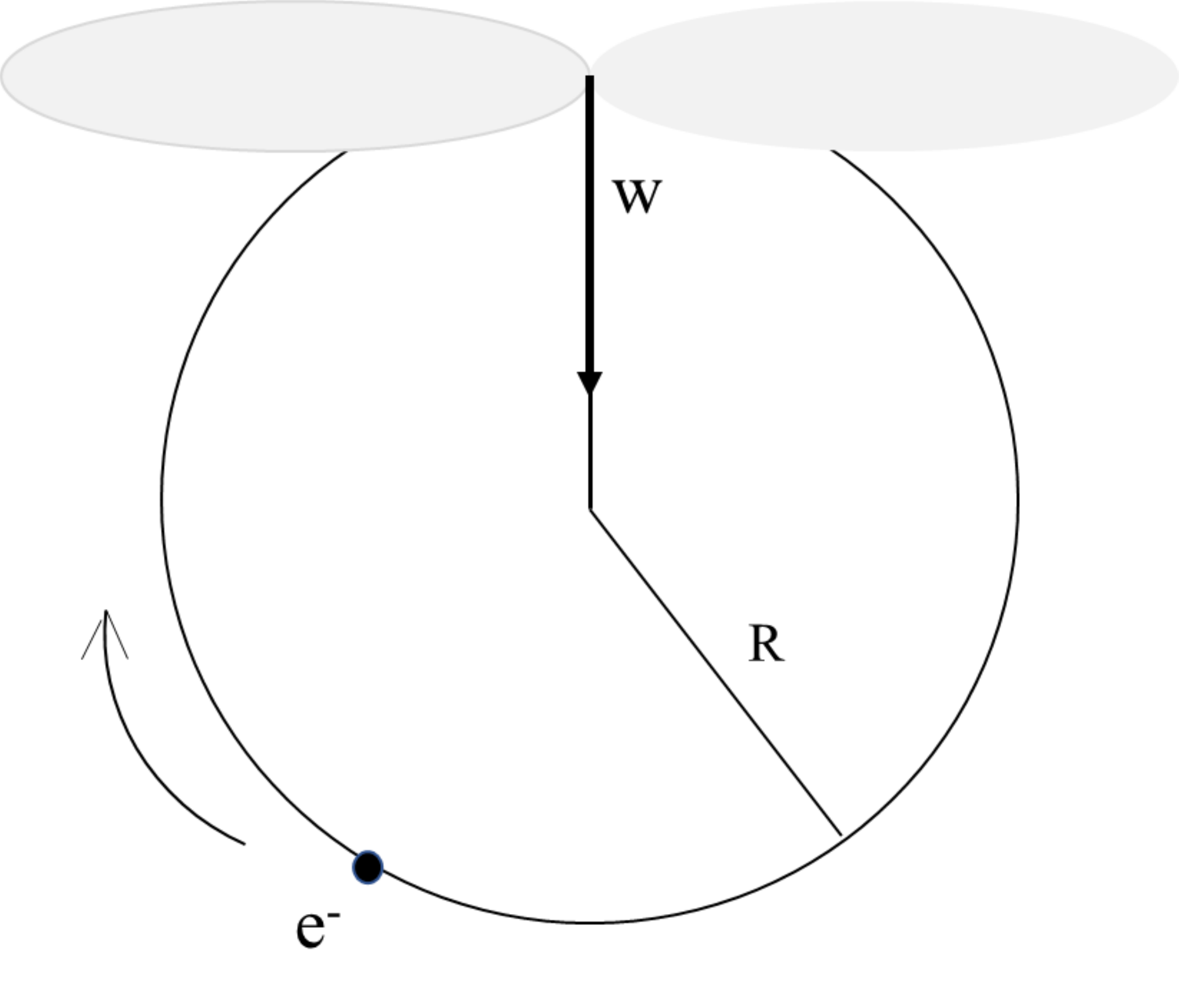}
\caption{The power radiated per solid angle by an accelerated charged particle is proportional to the angle $\theta$, which is the angle between the acceleration of the particle and the position vector of the observer: $dP/d\Omega\sim q^2w^2\sin^2{\theta}/c^3$.  This expression maximizes at $\theta=\pm90^\circ$ and produces the familiar figure $\infty$ shape that we are accustomed to associate, for example, with the emission pattern of the antenna of a radio station. \label{Fig:photon}}\end{figure}

Knowing that the transition occurs with the emission of a photon by an electron traveling in a circular orbit, we expect that the transition rate will involve both the charge of the electron $e$ and its acceleration $w$ . Note that if the particle was not accelerating, it would have been possible to find a different coordinate frame with respect to which the particle would have been at rest; therefore the relevant quantity for our analysis is the particle's acceleration and not its velocity (Fig. \ref{Fig:photon}).

Since the photon is traveling at the speed of light $c$, we also expect our final result to contain  $c$. In this case:
$$[e^2/LT]=[w]^\nu [e]^\beta[c]^\gamma=[L/T^2]^\nu[e]^\beta[L/T]^\gamma.$$
Equating the dimensions,
\begin{equation}\nonumber \quad -1=\nu+\gamma, \ 2=\beta , \ -1=-2\nu-\gamma , \end{equation}
yields:$$\nu=2, \beta=2, \gamma=-3, \quad {\rm i. e. }\quad
P\sim\frac{e^2w^2}{c^3}$$
To the accuracy of a numerical coefficient 2/3, this coincides with the Larmor formula in classical electrodynamics. \cite{jackson}.

If we assume that an electron is transitioning from a $2p$ state to an $1s$ state, the total radiated power will be equal to the difference in the energy of the two states times the time of the transition $\tau $.
 We can conclude that the time of the transition is: $$\tau^{-1}=\frac{P}{Ry}= \frac{e^2w^2}{c^3}\cdot\frac{\hbar^2}{m_e e^4}$$
or
\begin{equation}\tau^{-1}=\frac{e^{10} }{c^3}\frac{m_e}{\hbar^6}=\frac{m_e e^4}{\hbar^3}\Big(\frac{e^2}{\hbar c}\Big)^3.\label{eq:lifetime}\end{equation}
To the accuracy of a numerical factor, our result is the same as the one derived rigorously \cite{fermi}.

One conclusion common to Sections \ref{sec:soi} and \ref{sec:trara} is that while their subjects are traditionally considered intimately quantum, they can be well understood in the realm of classical physics when proper quantum parameters (energy levels, orbit radii, etc.) are used.

\section{Quantum electrodynamics}
\begin{figure*}[tb]
\includegraphics[width=0.60\textwidth]{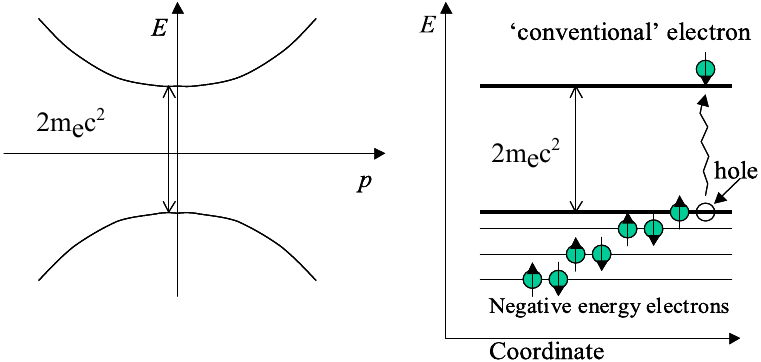}
\caption{The negative and positive branches of the dispersion Eq. (\ref{eq:dirac}) corresponding to the negative
and positive electron energies (left) and the corresponding energy diagram in real space (right) where the hole
corresponds to the effectively positive particle called positron. \label{Fig:dirac}}\end{figure*}
The famous energy-momentum equation derived by Einstein: $E^2=(mc^2)^2+(pc)^2$ was reinterpreted by Dirac as
\begin{equation}
E=\pm\sqrt{(mc^2)^2+(pc)^2}.\label{eq:dirac}\end{equation} Dirac suggested that the negative sign branch is {\it fully} occupied by negative energy particles (e. g. electrons) which however cannot
contribute to any physical effects, since there is no room for them to move (all the states are already occupied;
Fig. \ref{Fig:dirac}). On the other hand, if an electron with originally negative energy is transferred into the continuum of the
`conventional' positive energy branch, then it becomes a `conventional' electron and can participate in many
physical phenomena. In addition, the `hole' left in the background of negative energy electrons will behave as a
{\it positively} charged quasiparticle, which is called positron.

The above understanding has many implications when blended into quantum mechanics. That blend forms the subject of quantum electrodynamics, some aspects of which we consider next.

\subsection{Compton length}\label{sec:cl}
\begin{figure}[tb]
\includegraphics[width=0.35\textwidth]{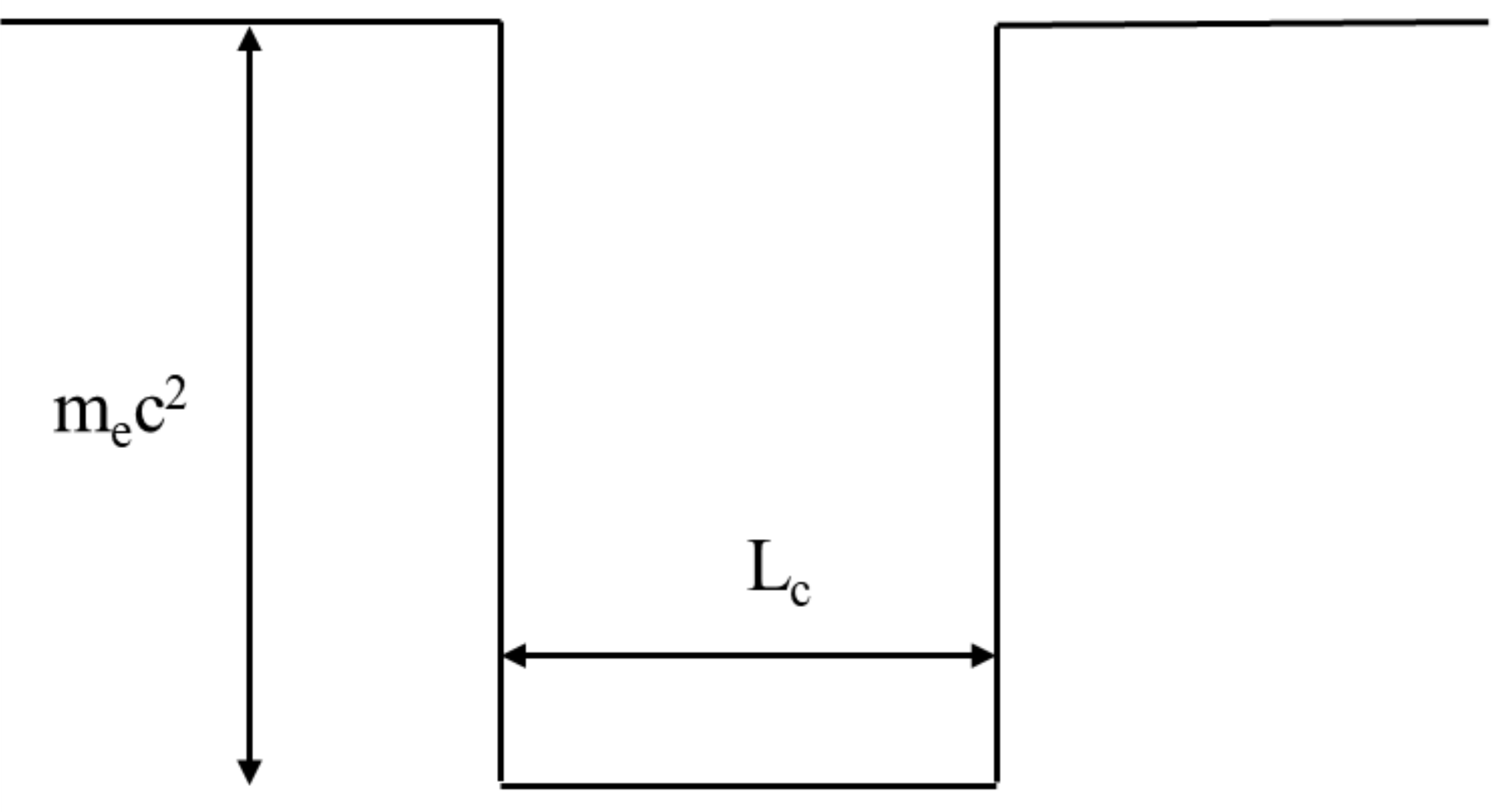}
\caption{Definition of the Compton length.\label{Fig:compton}}\end{figure}

Our first step will reconcile the two band particle-antiparticle structure with the uncertainty principle, which according to Eq. (\ref{eq:kine}) predicts that the kinetic energy increases as the particle localization length decreases. In the order of magnitude, the minimum localization length is determined by the condition that the kinetic energy increase $\hbar ^2/m_eL^2$ becomes comparable to the gap width $m_e c^2$ as illustrated in Fig. \ref{Fig:compton}. The latter criterion yields the value
\begin{equation}\label{eq:com}L_c = \frac{\hbar}{m_e c}\end{equation}
that is known as the Compton length.

Migdal \cite{migdal} assumed that when the electric field strength $F$ becomes high enough to increase the electron energy by $m_e c^2$ across the distance $L_c$ given by Eq. (\ref{eq:com}), then the vacuum state becomes unstable since the electrons with negative energies move to the upper band. His condition, $eFL_c\sim m_e c^2$, predicts the maximum field strength above which the vacuum breaks and the Maxwell equations are no longer applicable (because of electric charges generated by the breakdown). That field strength is:
\begin{equation}\label{eq:vacbreak}F\sim\frac{m_e^2c^3}{e\hbar}\sim\frac{e^2}{(\hbar ^2/m_e e^2)^2}\left(\frac{\hbar c}{e^2}\right)^3=F_{at}\frac{1}{\alpha ^3}\end{equation}
where $F_{at}=e^2/a_0^2$ is the characteristic atomic electric field, $a_0=\hbar ^2/m_e e^2$ is
the characteristic (Bohr) radius, and $\alpha\approx 1/137$ is the fine structure parameter. We conclude that classical electrodynamics remains adequate for electric fields with strength of a million times smaller than that of the atomic electric field.

While the above criterion $eFL_c\sim m_e c^2$ appears intuitively conceivable, it leaves multiple questions unanswered, such as why the minimum localization length is used in this case, an issue that will be clarified in the next section.

%

\subsection{Breakdown by tunneling}\label{sec:bd}
\begin{figure}[tb]
\includegraphics[width=0.20\textwidth]{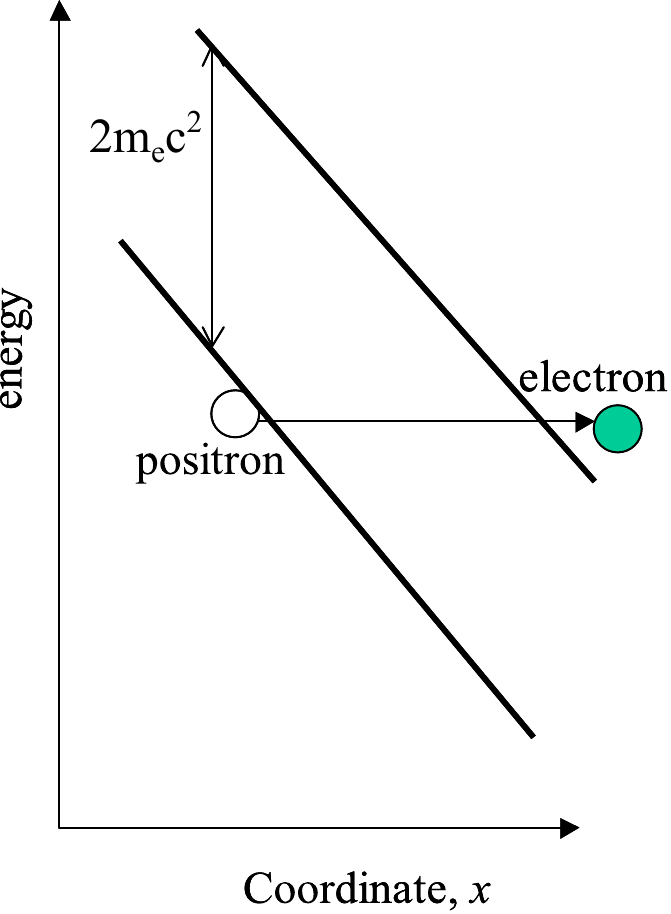}
\caption{Spontaneous electron-positron pair creation violating the classical electrodynamics Maxwell equations in
strong electric field.\label{Fig:ehpair}}\end{figure}

 One of the unanswered questions that was mentioned in the previous section is the violation of the
classical electrodynamics' Maxwell equations due to spontaneous electron-positron pair generation in a strong
enough electric field as illustrated in Fig. \ref{Fig:ehpair}. The generated pairs are not accounted for in the Maxwell
equations and will lead to effects beyond classical electrodynamics. Here, we will estimate the characteristic
field strength above which this violation takes place.

The required estimate is based on the observation that the pair creation occurs through the tunneling under the
barrier of characteristic height $V=2m_ec^2$ and width $a=V/Fe$ where $F$ is the electric field strength. The
probability of tunneling is
$$P\sim\exp \left(-2a\sqrt{\frac{2m_e V}{\hbar ^2}}\right)=\exp \left(-\frac{4m_e c^2}{Fe}\sqrt{\frac{2m_e ^2c^2}{\hbar ^2}}\right).$$
The process of pair creation becomes important when $P$ is not too small, i. e. the exponent in the latter equation is roughly of the order of the unity. The latter criterion leads directly to the estimate in Eq. (\ref{eq:vacbreak}).

In addition, from the above consideration, it is straightforward to derive that the tunneling length (horizontal arrow in Fig. \ref{Fig:ehpair}) equals  the Compton length from Eq. (\ref{eq:com}), which answers the question raised at the end of the previous section.

%

\section{Conclusions}
We have tried to present sequential examples  that form certain patterns potentially recognizable from other problems -not addressed here- such as dimensions, quantization, tunneling, etc. The information-overwhelming environment obligates us to hone our  abilities in pattern-recognition, since the available amount of facts and artifacts is increasing exponentially on a daily basis. We note however that the inclusion of qualitative methods is not meant as a complete substitute for a traditional, well-organized undergraduate or graduate QM class where mathematical derivations follow conceptual content.

The above set of examples appears accessible to college and grad students and can be blended indeed into the existing QM curricula. We have successfully used those examples teaching 4000 and 7000 level courses, and have observed  higher enthusiasm towards both in-class and home assignments compared to the standard QM items. We attribute that positive feedback to an appreciation for the shorter learning curve provided by the qualitative methods along with their inductive approach being natural to our cognition. The injection of qualitative methods will arm not only students, but acting professionals as well, with the toolkit needed as they deal with the new realities.

Along with qualitative tools typical of other fields in physics, such as dimensional analyses and various approximations, a more specific aspect related to QM is the original `naive' De Broglie approach typically neglected in the existing curricula in favor of the Schrodinger equation. That skew appears rather detrimental; it would be similar to the hardly imaginable scenario of geometrical optics being neglected in favor of wave analysis (note that the WKB approximation, while conceptually close to the De Broglie approach and included in the standard QM curricula, remains rather formal and mathematically involved). We hope that our examples in Sec. \ref{sec:DBQM} will offer enough motivation to present more of the De Broglie QM.

Multiple other topics related to qualitative methods in QM not addressed here could include the macroscopic quantum phenomena: superfluidity, superconductivity, and quantum Hall effect. They can perhaps better fit other possible compendia, such as the qualitative methods in condensed matter physics, to be presented elsewhere.

\section*{Acknowledgements}
We are grateful to J. G. Amar,  B. Gao, S. V. Khare, N. H. Redington,  A. V. Subashiev and V. I. Tsidilkovskii for reading our manuscript and useful comments.

\end{document}